\documentstyle[sprocl,epsfig]{article}

\def\st{\ifmmode{\tilde{t}} \else{$\tilde{t}$} \fi}
\def\sb{\ifmmode{\tilde{b}} \else{$\tilde{b}$} \fi}
\def\sq{\ifmmode{\tilde{q}} \else{$\tilde{q}$} \fi}
\def\sg{\ifmmode{\tilde{g}} \else{$\tilde{g}$} \fi}
\def\sz{\ifmmode{\tilde{\chi}^0} \else{$\tilde{\chi}^0$} \fi}
\def\sw{\ifmmode{\tilde{\chi}} \else{$\tilde{\chi}$} \fi}
\def\sl{\ifmmode{\tilde{\ell}} \else{$\tilde{\ell}$} \fi}
\def\sn{\ifmmode{\tilde{\nu}} \else{$\tilde{\nu}$} \fi}
\def\stau{\ifmmode{\tilde{\tau}} \else{$\tilde{\tau}$} \fi}
\def\nle{\rlap{\lower 3.5 pt 
\hbox{$\mathchar \sim$}}\raise 1pt \hbox{$<$}}
\def\nge{\rlap{\lower 3.5 pt 
\hbox{$\mathchar \sim$}}\raise 1pt \hbox{$>$}}

\begin{document}

\vspace*{-2.5cm}
\begin{flushright}
  UWThPh-1997-47 \\
  HEPHY-PUB 680/97 \\
  hep-ph/9712214
\end{flushright}
\vspace*{0.6cm}

\title{SUSY--QCD Corrections to \\
       Higgs Particle Decays into Quarks and Squarks
      }

\author{\underline{A.~Bartl}$^{1)\dagger}$, H.~Eberl$^{2)}$, K.~Hidaka$^{3)}$,
        T.~Kon$^{4)}$, W.~Majerotto$^{2)}$, Y.~Yamada$^{5)}$
       }

\address{
$^{1)}$ Institut f\"ur Theoretische Physik, Universit\"at Wien,
A-1090 Vienna, Austria\\
$^{2)}$ Institut f\"ur Hochenergiephysik der \"Osterreichischen
Akademie der Wissenschaften,  A-1050 Vienna, Austria\\
$^{3)}$ Department of Physics, Tokyo Gakugei University, Koganei,
Tokyo 184, Japan\\
$^{4)}$ Faculty of Engineering, Seikei University, Musashino,
Tokyo 180, Japan\\
$^{5)}$ Department of Physics, Tohoku University, Sendai 980-77, Japan
         }

\maketitle
\begin{abstract}
We study the decays of the Higgs bosons $H^{\pm}$,
$H^0$, and $A^0$ within the Minimal Supersymmetric Standard
Model. For decays into quarks and squarks we include the
supersymmetric QCD radiative corrections. We find that the
corrections are significant and can go up to 50\%. The
supersymmetric decay modes 
$H^+ \rightarrow \tilde t \bar{\tilde b}$ and 
$H^0 (A^0) \rightarrow \tilde t \bar{\tilde t}$ can be dominant
in a wide range of the model parameters due to the large Yukawa couplings
and mixings of $\tilde t$ and $\tilde b$.
\end{abstract}

\footnotetext{$\dagger$ Talk presented at the
{\em International Workshop on Quantum Effects in the MSSM},
September 9 -- 13, 1997, Barcelona, Spain.}
 
\section{Introduction}
We need to find the Higgs boson for a conclusive test of the
electroweak symmetry breaking mechanism
of the Standard Model. The
search for the Higgs boson, therefore, has high priority at LEP,
TEVATRON, LHC, and a future $e^+e^-$ Linear Collider. 
To facilitate searching for the Higgs boson we need to
study not only the production mechanisms, but also all possible
decay modes. While in
the Standard Model (SM) there is only one physical Higgs
particle, extensions of the SM contain more Higgs
bosons. 

In this contribution we consider Higgs particle decays in
the Minimal Supersymmetric Standard Model (MSSM)
\cite{bartl:cit1}. The MSSM implies the existence of 
five physical Higgs bosons $h^0$, $H^0$, $A^0$, and $H^\pm$
\cite{bartl:cit2,bartl:cit3}.
Provided that all SUSY particles are very heavy, the $H^+$ decays 
mainly into $t\bar{b}$, and below the $t\bar{b}$ threshold
the decays $H^+ \to \tau^+ \nu$ and/or $H^+ \to W^+ h^0$ are
dominant  \cite{bartl:cit2,bartl:cit4}.
Similarly, if all decay modes of $H^0$ and $A^0$ into SUSY
particles are kinematically forbidden, they decay dominantly into a
fermion pair of the third generation.
Higgs boson decays into supersymmetric (SUSY) particles 
can be very important if they are kinematically allowed.
The decays into charginos and neutralinos can have
large branching ratios, and can significantly change the
signatures of SUSY Higgs particles \cite{bartl:GuHa,bartl:DjOhm}. 
The decays into squarks can be the dominant decay modes of Higgs
bosons in a large parameter region in case that the squarks
are relatively light
\cite{bartl:citH+tree,bartl:citH0tree}.

For a precise determination of the Higgs boson couplings to
quarks and squarks we need to include the SUSY--QCD
corrections in the calculation of the decay widths.
The SUSY--QCD corrections in ${\cal O}(\alpha_s)$ were calculated in the
on--shell scheme for the
processes $H^+ \to t \bar b$ in \cite{bartl:citH+qcd,bartl:Jimenez},
and for $H^0, A^0 \to q \bar{q}$
in \cite{bartl:Coarasa}. For the decays of
Higgs particles into squark pairs we calculated the SUSY--QCD corrections
in the on--shell scheme in \cite{bartl:citHiggsqcd}, including
squark--mixing and a proper renormalization of the mixing angle
$\theta_{\tilde q}$ \cite{bartl:squarkprod}. The SUSY--QCD
corrections to Higgs boson decays into squarks were also studied
in \cite{bartl:hollik} recently.

In this talk we review our work on the branching ratios of Higgs
boson decays. In order to show how the branching ratios of the
various decay modes depend on the SUSY parameters, we will first 
summarize the tree--level results. Then we will take into account
the SUSY--QCD corrections in ${\cal O}(\alpha_s)$ for the decay
branching ratios into third generation quarks and squarks. We will show
that in most cases the SUSY--QCD corrections are significant and
need to be included.

At tree--level the
masses of the MSSM Higgs bosons depend on the two parameters $m_A$
and  $\tan\beta$. $m_A$ is
the mass of the pseudoscalar Higgs boson $A^0$, and
 $\tan\beta=\frac{v_2}{v_1}$ is the ratio of the vacuum
expectation values of the two neutral Higgs doublet states
\cite{bartl:cit2,bartl:cit3}.
The mass of $h^0$ gets
large radiative corrections from one--loop contributions.
We will take into account these corrections using the formulae 
of \cite{bartl:radhiggs}.
The experimental lower bounds on the Higgs boson masses from LEP 
are $m_{h^0} > 62$~GeV and $m_{A^0} > 62$~GeV \cite{bartl:aleph}.
In addition to $\tan\beta$, the main SUSY parameters in the chargino 
and neutralino systems are the Higgs--higgsino mass parameter 
$\mu$ and the $SU(2)$ gaugino mass parameter $M$. We assume that
$M$ is related to the gluino mass $m_{\sg}$ and the $U(1)$
gaugino mass parameter $M'$ by
$M=(\alpha_2/\alpha_s(m_\sg))m_{\sg}=3/(5\tan^2\theta_W)M'$.
For the third generation squarks and sleptons we also need 
the mass parameters $M_{\tilde{Q}}$, $M_{\tilde{U}}$, 
$M_{\tilde{D}}$, $M_{\tilde{L}}$, $M_{\tilde{E}}$,
and the trilinear scalar coupling parameters $A_t$, $A_b$ and
$A_{\tau}$. 

\section {Tree--Level Widths}
In the following we will use the short--hand notation
$H^k$, $k=1, 2, 3, 4$, for the Higgs bosons of the MSSM, with 
$H^1 \equiv h^0$,
$H^2 \equiv H^0$,
$H^3 \equiv A^0$, and $H^4 \equiv H^+$.
The decay widths for $H^k\rightarrow q \bar q$, $k=1, 2, 3$,
$q=t, b$, and $H^+\rightarrow t \bar b$ 
are given by
 \begin{eqnarray}
\Gamma^{tree}(H^k\rightarrow q \bar q)  & = & \frac{3 g^2 m_q^2
(d_k^q)^2 m_{H^k}}{32 \pi m_W^2 \, a_q^2}\left(1 - \frac{4
m_q^2}{m_{H^k}^2}\right)^{(3/2 - \delta_{k3})}\!\!\!\!\!\!\!\! (k = 1,2,3),
                         \\
\Gamma^{tree}(H^+\rightarrow t \bar b)  & = &
\frac{3 \kappa(m_{H^+}^2, m_t^2, m_b^2)}{16 \pi m_{H^+}^3} \times
    \nonumber\\
&& \left[
(m_{H^+}^2 - m_t^2 - m_b^2) (y_t^2 + y_b^2) - 4 m_t m_b y_t y_b\right]\, ,
  \label{bartl:H+quarks}
 \end{eqnarray}
with $\kappa(x,y,z)\equiv ((x-y-z)^2-4yz)^{1/2}$,
$d_1^t =  d_2^b = \cos\alpha$, $d_2^t =  -d_1^b = -\sin\alpha$,
$d_3^t = a_b = \cos\beta$, 
$d_3^b = -a_t = -\sin\beta$, where
$\alpha$ is the mixing angle in the $h^0$ -- $H^0$ system
\cite{bartl:cit2,bartl:cit3}. $y_t$ and $y_b$ are related to the 
Yukawa couplings and are
  $y_t = h_t \cos\beta = g m_t \cot\beta /(\sqrt{2} m_W)$ and
  $y_b = h_b \sin\beta = g m_b \tan\beta /(\sqrt{2} m_W)$.
The decay widths into top quarks are large due to the large top
quark mass. If $\tan\beta > 20$, the decay modes into
bottom quarks can also become important. Asymptotically for
$m_{H^k} \gg m_q$ the Higgs boson decay widths into quarks are
proportional to $m_{H^k}$.

The Higgs boson decay widths into squarks of the third
generation depend on $\sq_L - \sq_R$ mixing. This
mixing is described by the squark mass matrix which in
the basis ($\sq_L$, $\sq_R$), $\sq=\st$ or $\sb$, and in the
diagonalized form is
\begin{equation} \label{bartl:eq1}
\left( \begin{array}{cc}m_{LL}^2 & m_{LR}^2 \\ m_{RL}^2 & m_{RR}^2
\end{array} \right)=
(R^{\sq})^{\dagger}\left( \begin{array}{cc}m_{\sq_1}^2 & 0 \\ 0 &
m_{\sq_2}^2 \end{array}
\right)R^{\sq}\, ,
\end{equation}
where $R^{\sq}_{i\alpha}$ is a $2 \times 2$ rotation matrix with
rotation angle $\theta_\sq$, and
\begin{eqnarray}
m_{LL}^2 &=& M_{\tilde{Q}}^2+m_q^2+m_Z^2\cos 2\beta
(I^{3L}_q-e_q\sin^2\theta_W)\, , \\
m_{RR}^2 &=& M_{\{\tilde{U},
\tilde{D}\}}^2+m_q^2+m_Z^2\cos 2\beta\, e_q 
\sin^2\theta_W\, , \\
m_{LR}^2 &=& m_{RL}^2 =
m_q(A_q-\mu(\tan\beta)^{-2I^{3L}_q}) \, . \label{bartl:eq4}
\end{eqnarray}
$I^{3L}_q$ and $e_q$ are the third component of isospin and the
electric charge of the quark $q$, and $\theta_W$ is the Weinberg
angle. 
The mass eigenstates $\sq_i, i=1, 2$, ($m_{\sq_1}<m_{\sq_2}$) are
related to the states $\sq_{\alpha}, \alpha=L,R$, by
$\sq_i=R^{\sq}_{i\alpha}\sq_{\alpha}$.

\noindent The widths of the decays 
$H^k \rightarrow \sq_i \bar{\sq}_j$ at tree--level are
\begin{equation} \label{bartl:eq6}
\Gamma^{tree}(H^k\rightarrow\sq_i\bar{\sq}_j) =
\frac{3 \kappa(m_{H^k}^2, m_{\sq_i}^2, m_{\sq_j}^2 )}{16\pi
m_{H^k}^3}|G_{ijk}^\sq|^2\,.
\end{equation}
For $k = 1, 2, 3$ we have $\sq = \st, \sb$,
and for $k = 4$ we have $\sq_i \equiv \st_i$,
$\sq_j \equiv \sb_j$, $(i,j = 1,2)$. 
The expressions for the couplings $G_{ijk}^\sq$
are given in \cite{bartl:citHiggsqcd}.
The Higgs boson decay widths into squarks can be large in the case of
large squark mixing. For example, the width of 
$A^0 \rightarrow \sq_1 \bar{\sq}_2$
is directly proportional to 
$|m_q (A_q (\tan\beta)^{-2I^{3L}_q} + \mu)|^2$. The same
expressions appear in the couplings $G_{ij4}$ for the decays
$H^+\rightarrow\st_i\bar{\sb}_j$. The $H^0\st_i \st_j$ 
couplings
can be large since they contain terms proportional to $m_t$, and
the $H^0\sb_i \sb_j$ couplings can be large if $\tan\beta > 20$. 
More details can be found in \cite{bartl:citH0tree}. For
$m_{H^k} \gg m_{\sq_i}$ the widths of Higgs boson decays into
squarks behave asymptotically like $1/m_{H^k}$.

In the calculation of the corresponding branching ratios we have
included the widths of the following $H^+$,
$H^0$ and $A^0$ decay modes:\\
\begin{tabular}{lrcl}
({\romannumeral 1}) &
$H^+$ &$\to$&
$t \bar b$, $c \bar s$, $\tau^+ \nu_\tau$, $W^+ h^0$, $\st_i \bar\sb_j$,
$\tilde{\chi}^+_k \tilde{\chi}^0_l$,  $\tilde{\tau}_i^+
\tilde{\nu}_\tau$, $\tilde{\ell}_L^+ \tilde{\nu}_\ell \,(\ell = e, \mu)$,\\
({\romannumeral 2})&
$H^0$ &$\to$& $t\bar{t}$, $b\bar{b}$, $c\bar{c}$,
$\tau^-\tau^+$, $W^+W^-$, $Z^0Z^0$, $h^0h^0$, $A^0 A^0$,
$W^\pm H^\mp$, \\
&&& $Z^0A^0$, $\st_i\bar{\st}_j$,
$\sb_i\bar{\sb}_j$, $\sl^-_i\sl^+_j$, $\sn_\ell\bar{\sn}_\ell$
($\ell = e, \mu, \tau$), $\sw^+_i\sw^-_j$,
$\sz_k\sz_l$, and\\
({\romannumeral 3})&
$A^0$ &$\to$& $t\bar{t}$, $b\bar{b}$, $c\bar{c}$,
$\tau^-\tau^+$, $Z^0h^0$, $\st_1\bar{\st}_2$, $\st_2\bar{\st}_1$,
$\sb_1\bar{\sb}_2$, $\sb_2\bar{\sb}_1$,
$\stau^-_1\stau^+_2$, $\stau^-_2\stau^+_1$,\\
&&& $\sw^+_i\sw^-_j$, $\sz_k\sz_l$.
\end{tabular}

\vspace{2mm}

Formulae for these widths are found e.~g. in \cite{bartl:cit2}.
We have not taken into account loop induced decay modes like
$H^+ \to W^+ Z^0, W^+ \gamma$, $H^0 \to gg, \gamma\gamma$ etc., 
and three-body decay modes \cite{bartl:gunion,bartl:zerwas1}.

We have shown in \cite{bartl:citH+tree,bartl:citH0tree} that the
branching ratios for
decays into squarks, $H^+ \to \st_i \bar{\sb}_j$, 
$H^0 \to \st_i \bar{\st}_j, \sb_i \bar{\sb}_j$, and 
$A^0 \to \st_1 \bar{\st}_2 ,\sb_1 \bar{\sb}_2$ can be
larger than $50\%$ in a sizeable region of the SUSY parameter space.
To illustrate this we show in Figs.~\ref{bartl:figII}a, and c the 
tree level decay widths (dashed lines)
$\Gamma (H^0 \to \st \bar{\st}) \equiv
\sum_{i,j} \Gamma (H^0 \to \st_i \bar{\st}_j)$ and 
$\Gamma (A^0 \to \st_1 \bar{\st}_2)$ as a function of $m_A$. For
comparison we also show in Figs.~\ref{bartl:figII}a and c 
the tree level decay
widths $\Gamma (H^0 \to t \bar{t})$ and
$\Gamma (A^0 \to t \bar{t})$ (dashed lines). Fig.~\ref{bartl:figIII}a 
shows the
tree level decay widths
$\Gamma (H^+ \to \st \bar{\sb}) \equiv
\sum_{i,j} \Gamma (H^+ \to \st_i \bar{\sb}_j)$ and
$\Gamma (H^+ \to t \bar{b})$ as a function of $m_A$ (dashed lines).
 In these plots we have assumed the relations
$M_{\tilde{Q}} : M_{\tilde{U}} : M_{\tilde{D}} 
 = 1 : \frac 8 9 : \frac{10}{9}$ for the
squark mass parameters, and
$A \equiv A_t = A_b$ for the trilinear scalar coupling
parameters. We have taken $M_{\tilde Q} =120$~GeV, 
$A=280$~GeV, $\mu=300$~GeV, $M=140$~GeV, and $\tan\beta
=3$. In the plots we have required $m_{h^0} > 70$~GeV.
For large $m_{A^0}$ one has $m_{H^+} \approx m_{H^0} \approx m_{A^0}$.
For this set of parameters we have (in GeV)
($m_{\st_1}$, $m_{\st_2}$, $m_{\sb_1}$, $m_{\sb_2}$, $m_{\sg}$,
$m_{\tilde \chi_1^0}$, $m_{\tilde \chi_1^+}$) =
(102, 271, 121, 145, 412, 63, 116). 
In the examples shown, the decay widths 
$\Gamma (H^0 \to \st \bar{\st})$,
$\Gamma (A^0 \to \st_1 \bar{\st}_2)$, and
$\Gamma (H^+ \to \st \bar{\sb})$ are always much larger than the
decay widths $\Gamma (H^0 \to t \bar{t})$,
$\Gamma (A^0 \to t \bar{t})$, and
$\Gamma (H^+ \to t \bar{b})$, respectively. The corresponding
branching ratios for decays into third generation squarks 
are much larger than 50\% in the whole $m_A$ range shown
where these decays are kinematically allowed.
The branching ratios for the decays into sbottoms, 
$H^0 \to \sb_i \bar{\sb}_j$ and $A^0 \to \sb_1 \bar{\sb}_2$
turn out to be less than 3\% due to the low $\tan\beta $ value
considered. 

\section{SUSY--QCD Corrected Decay Widths}
In Section 2 we have seen that the Higgs boson decays into 
squarks can be important. Therefore, it is
necessary to include the SUSY--QCD radiative corrections in the
calculation of the widths for the decays into squarks and into
quarks. In this section we will review some of our results about
the branching ratios of Higgs boson decays including 
the SUSY--QCD corrections in ${\cal O}(\alpha_s)$. For
further details concerning the theoretical calculation of these 
corrections we refer to 
\cite{bartl:citH+qcd,bartl:Jimenez,bartl:Coarasa,bartl:citHiggsqcd,bartl:hollik}.
 
\begin{figure}
\begin{center}
\hspace*{-4mm}\mbox{\epsfig{figure=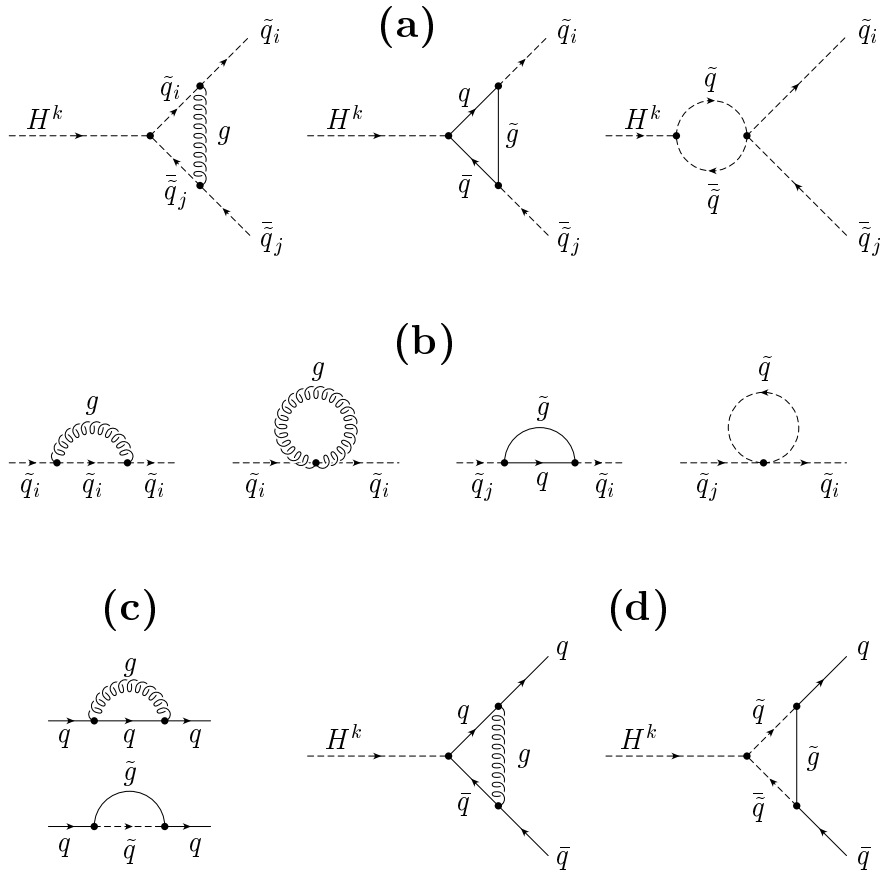,width=12cm}}
\end{center}
\caption[bartl:fig1]{Feynman diagrams for the 
calculation of the virtual ${\cal O}(\alpha_s)$ SUSY--QCD
corrections to the decay widths $H^k \to \sq_i \bar\sq_j$ 
and $H^k \to q \bar q$. \label{bartl:figI}}
\end{figure}

The Feynman diagrams for the virtual ${\cal O}(\alpha_s)$ SUSY--QCD
corrections are shown in Fig.~\ref{bartl:figI}. 
We work in the on--shell
renormalization scheme. We first discuss the radiative
corrections for the Higgs boson decays into squarks. In this
case the virtual corrections consist of
the vertex corrections, wave function corrections, and
the corrections due to the shift from the bare couplings to the
on--shell couplings. We use the scheme
introduced in \cite{bartl:squarkprod} for 
$e^+e^- \to \sq_i \bar\sq_j$, where we fixed the counterterm
of the squark mixing angle such that it cancels the
off--diagonal term of the squark wave--function correction to
$e^+e^- \to \sq_1 \bar\sq_2$. For the shift 
$\delta \theta_{\tilde q}$ we take the same
expression as in \cite{bartl:squarkprod}. A more detailed
discussion of the on--shell renormalization of the squark mixing
angle $\theta_{\tilde q}$ is given in \cite{bartl:majer}.

%
The calculation of the SUSY--QCD corrections to the decay widths
of $H^+ \to \st \bar{\sb}$ and of the branching ratios of $H^0$
and $A^0$ decays involves both the stop 
and the sbottom sector. We have to pay special attention to the
parameter $M_{\tilde Q}$ in the on--shell scheme. $SU(2)_L$
symmetry requires that at tree--level and in the $\overline{DR}$
scheme the parameter $M_{\tilde Q}$ is the same in the stop and 
sbottom mass matrix (see eq.(4)). 
However, in the on--shell scheme this is no more the
case, because the shifts from the $\overline{DR}$ parameters to the
on--shell parameters are, in general, different for the stop 
and sbottom sectors.
In the present case we choose $M_{\tilde Q}(\st )|_{OS}$ in
the stop sector as the on--shell input parameter. Then 
$M_{\tilde Q}(\sb )|_{OS}$ in the sbottom sector is shifted by
the amount
\begin{equation}
M^2_{\tilde Q}(\sb )|_{OS} = M^2_{\tilde Q}(\st )|_{OS} +
\delta M^2_{\tilde Q}(\st ) -  \delta M^2_{\tilde Q}(\sb) \, .
\label{bartl:eq8}
\end{equation}
The  shift 
$\delta M^2_{\tilde Q}(\st ) -  \delta M^2_{\tilde Q}(\sb )$ is
ultra--violet finite due to the underlying $SU(2)_L$ symmetry.

We also include the SUSY--QCD corrections for the Higgs decays
into third generation quarks, taking the formulae of
 \cite{bartl:citH+qcd,bartl:Jimenez,bartl:Coarasa}.
In the following numerical examples we assume for the
on--shell input parameters the same relations as for the
tree--level quantities, 
$M_{\tilde{Q}}(\st) : M_{\tilde{U}} : M_{\tilde{D}} : M_{\tilde{L}} :
M_{\tilde{E}} = 1 : \frac 8 9 : \frac{10}{9} : 1 : 1$ and
$A \equiv A_t = A_b = A_\tau$.
We take $m_t = 175$~GeV, $m_b = 5$~GeV, $m_Z = 91.2$~GeV, $m_W = 80$~GeV,
$\sin^2\theta_W=0.23$, $\alpha_2 = 0.0337$,
and $\alpha_s = \alpha_s(m_{H^k})$ for $H^k$ decay. We use
$\alpha_s(Q)=
12\pi/\{(33-2n_f)\ln(Q^2/\Lambda_{n_f}^2)\}$,
with $\alpha_s(m_Z)=0.12$, and the number of quark flavors $n_f=5(6)$ for
$m_b<Q\le m_t$ (for $Q>m_t$).

\begin{figure}
\begin{center}
\mbox{\epsfig{figure=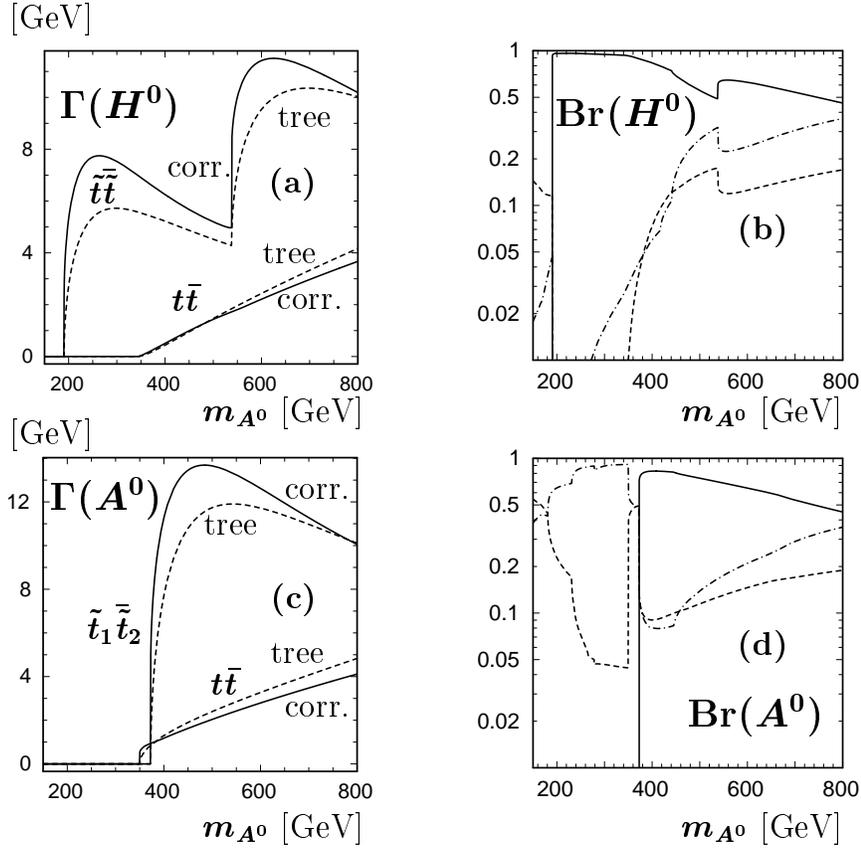,width=12cm}}
\end{center}
\caption[bartl:fig2]{Tree--level and SUSY--QCD corrected decay
widths into squarks and quarks
((a) and (c)) and important branching ratios ((b) and (d))
for the neutral Higgs boson decays,
$H^0, A^0 \to \sum_{i,j=1,2}(\st_i \bar\st_j + \sb_i \bar\sb_j)$
(full line), $H^0, A^0 \to t \bar t + b \bar b$ (dashed line),
and $H^0, A^0 \to \sum_{i,j=1,2} \tilde\chi_i^+  \tilde\chi_j^- +
\sum_{i,j=1,4} \tilde\chi_i^0  \tilde\chi_j^0$ (dashed--dotted line),
as functions of $m_{A^0}$. \label{bartl:figII}}
\end{figure}

\begin{figure}
\begin{center}
\mbox{\epsfig{figure=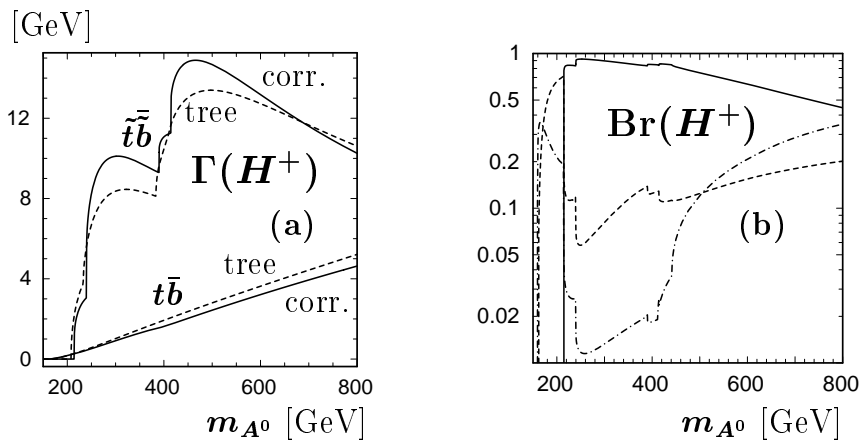,width=12cm}}
\end{center}
\caption[bartl:fig3]{(a) tree--level and SUSY--QCD corrected decay
widths into squarks and quarks
and (b) important branching ratios 
for the charged Higgs boson decays,
$H^+ \to \sum_{i,j=1,2} \st_i \bar\sb_j$
(full line), $H^+ \to t \bar b$ (dashed line),
and $H^+ \to \sum_{i=1,2} \sum_{j=1,4}  \tilde\chi_i^+  \tilde\chi_j^0$
(dashed--dotted line), as functions of $m_{A^0}$. \label{bartl:figIII}}
\end{figure}

In addition to the tree--level decay width we show in 
Fig.~\ref{bartl:figII}a
also the SUSY--QCD corrected decay width
$\sum_{i,j} \Gamma(H^0 \to  \st_i \bar{\st}_j)$ and
$\Gamma(H^0 \to  t \bar{t})$ as a function of $m_A$ (full lines).
In Fig.~\ref{bartl:figII}c we show also the SUSY--QCD corrected widths
$\Gamma(A^0 \to  \st_1 \bar{\st}_2)$ and
$\Gamma(A^0 \to  t \bar{t})$, and in Fig.~\ref{bartl:figIII}a those of
$\sum_{i,j} \Gamma(H^+ \to  \st_i \bar{\sb}_j)$ and
$\Gamma(H^+ \to  t \bar{b})$. We have taken 
$M_{\tilde{Q}}(\st) = 120$~GeV, and for $A, M, \mu$, and
$\tan\beta$  the same values as in the tree--level calculation.
The masses of $\st_1$, $\st_2$, $\sg$, $\tilde \chi_1^0$, and
$\tilde \chi_1^+$ are the same as mentioned at the end of Section~2,
however, those of 
$\sb_1$ and $\sb_2$ are different due to
eq.~(\ref{bartl:eq8}). For the parameters used we get
$M_{\tilde{Q}}(\sb) = 134$~GeV, $m_{\sb_1} = 127$~GeV, and
$m_{\sb_2} = 151$~GeV.    
This means that the shift 
$(M_{\tilde Q}(\sb) - M_{\tilde Q}(\st))|_{OS}$ at one--loop
level is about 10\%
of the tree--level value of $M_{\tilde Q}$. As can be seen in
Figs.~\ref{bartl:figII}a,~c and Fig.~\ref{bartl:figIII}a, 
the corrections to the sums
of the decay widths
$\sum_{i,j} \Gamma(H^0 \to \st_i \bar\st_j)$,
$\sum_{i,j} \Gamma(H^+ \to \st_i \bar\sb_j)$, and to
$\Gamma(A^0 \to \st_1 \bar\st_2)$ are significant and can be
larger than 30\%.
The modes into bottom
quarks and sbottoms are very small compared to the top and stop modes 
and are not shown. In Figs.~\ref{bartl:figII}b and d 
we show the SUSY--QCD
corrected branching ratios for $H^0$ and $A^0$ decays into
squarks, quarks, charginos and neutralinos, 
and in Fig.~\ref{bartl:figIII}b
those for $H^+$ decays. In the examples shown the squark decay
modes are always the dominant ones.
The discontinuities in
$\sum_{i,j} \Gamma(H^0 \to \st_i \bar\st_j)$ and
$\sum_{i,j} \Gamma(H^+ \to \st_i \bar\sb_j)$, and in the
corresponding branching ratios, are due to decay channels opening.\\
The SUSY--QCD corrections to the widths of individual
decay modes into squarks, $H^0 \to \st_i \bar\st_j$ or
$H^+ \to \st_i \bar\sb_j$, may go up to 50\%. They may also
be negative. When summed over the individual decay channels, the
SUSY--QCD corrections to
$\sum_{i,j} \Gamma(H^0 \to \st_i \bar\st_j)$ and
$\sum_{i,j} \Gamma(H^+ \to \st_i \bar\sb_j)$ are in many cases
positive, whereas those for the decays into quarks
are in general negative.
Therefore, in these cases the branching ratios for decays into
squarks are enhanced by including the SUSY--QCD corrections.  

\begin{figure}
\begin{center}
\mbox{\epsfig{figure=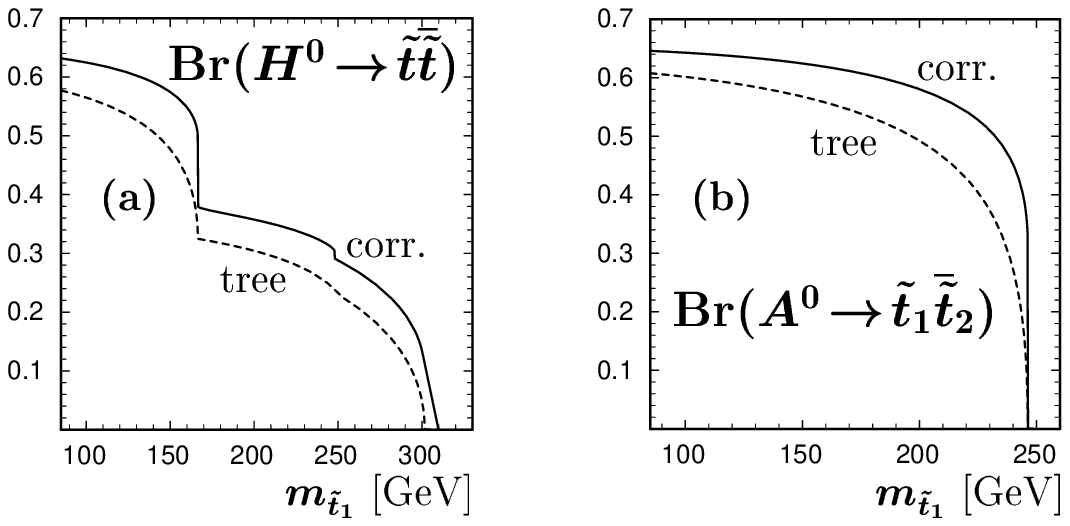,width=12cm}}
\end{center}
\caption[bartl:fig4]{Tree--level and SUSY--QCD corrected branching 
fractions of the neutral Higgs bosons $H^0$ and $A^0$
decaying into squarks, as
a function of $m_{\st_1}$. \label{bartl:figIV}}
\end{figure}

\begin{figure}[t]
\begin{center}
\mbox{\epsfig{figure=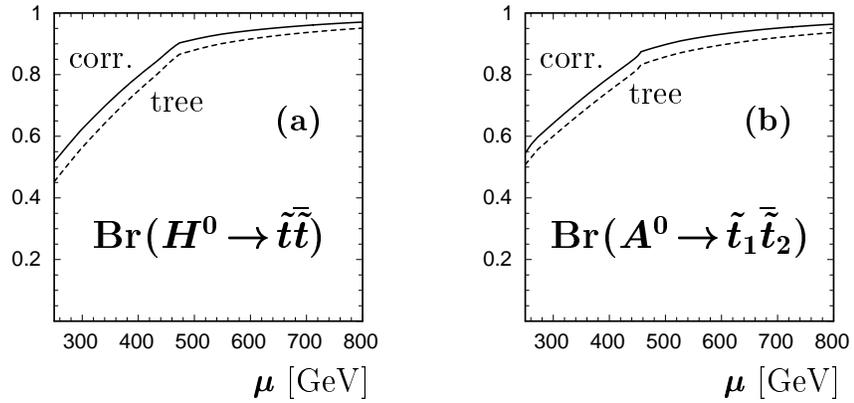,width=12cm}}
\end{center}
\caption[bartl:fig5]{Tree--level and SUSY--QCD corrected branching 
fractions of the neutral Higgs bosons $H^0$ and $A^0$ decaying
into squarks, as
functions of $\mu$. \label{bartl:figV}}
\end{figure}

We also studied the $m_{\tilde t_1}$ and $\mu $ dependence of the 
tree--level and SUSY--QCD corrected branching ratios. 
Figs.~\ref{bartl:figIV}a
and b show the branching ratios 
$\sum_{i,j} B(H^0 \to \st_i \bar{\st}_j)$ and
$B(A^0 \to \st_1 \bar{\st}_2)$ as a function of $m_{\tilde
t_1}$ by varying $M_{\tilde Q} = M_{\tilde Q}(\st)$,
taking $m_A = 600$~GeV, $\mu =300$~GeV, $M=140$~GeV,
$\tan\beta =3$, and $A=280$~GeV. Figs.~\ref{bartl:figV}a and b 
show the same
branching ratios as a function of $\mu $, taking $M_{\tilde Q} =
120$~GeV, and the remaining parameters as in Figs.~\ref{bartl:figIII}.
For $\mu < 500$~GeV the branching ratios for $H^0$ and $A^0$ decays into
squarks increase with increasing $\mu$. This is a consequence of the 
$\mu$--dependence of the widths for the decays into charginos and
neutralinos.

\section*{Acknowledgements}
We are very grateful to Prof. Joan Sol\`a for his kind invitation
to this interesting workshop. We really enjoyed its inspiring
atmosphere and its informative character.
This work was supported by the "Fonds zur F\"orderung der
wissenschaftlichen Forschung" of Austria, project no. P10843--PHY.


\section*{References}


\end{document}